\documentclass[notitlepage,twocolumn]{article}
\usepackage[left=1in, right=1in, top=1in, bottom=1in]{geometry}

\usepackage{titling}
\usepackage{multicol,lipsum}
\usepackage{authblk}
\usepackage{graphicx}
\pretitle{\begin{center}\LARGE\bfseries}
	\posttitle{\par\end{center}\vskip 0.5em}
\preauthor{\begin{center}\Large\ttfamily}
	\postauthor{\end{center}}
\predate{\par\large\centering}
\postdate{\par}

\title{In-situ observation of temperature-dependent segregation of Ni adatoms on oriented Pd surfaces}
\author{Samantha Zimnik$^1$, Marcel Dickmann$^1$ and Christoph Hugenschmidt}
\affil[1]{{Heinz Maier-Leibnitz Zentrum (MLZ) and Physik Department E21, Technische Universit{\"a}t M{\"u}nchen \newline Lichtenbergstra{\ss}e 1, 85748 Garching, Germany }}
\date{}                     
\newcommand{\dC}{\,$^\circ$C }
\begin{document}

	\thispagestyle{empty}
\twocolumn[
\begin{@twocolumnfalse}
	\maketitle
	\thispagestyle{empty}
	\begin{abstract}
		We report the direct observation of the in-situ temperature-dependent surface segregation of Ni adatoms on single crystalline Pd surfaces using Positron annihilation induced Auger Electron Spectroscopy (PAES). 
	For this study, a single atomic layer of Ni was grown on Pd with the crystallographic orientations Pd(111), Pd(110) and Pd(100). The sample temperature was increased from room temperature to 350\,$^\circ$C and the intensity of the Ni and Pd signal was evaluated from the recorded PAES spectra.
	Due to the outstanding surface sensitivity of PAES a clear tendency for Pd segregation was observed for all samples.  
	Moreover the activation temperature T$_0$ for surface segregation was found to depend strongly on the surface orientation: We determined T$_0$ to 172$\pm$4\,$^\circ$C, 261$\pm$12\,$^\circ$C and 326$\pm$11\,$^\circ$C for Pd(111), Pd(100) and Pd(110), respectively.\newline 
	\end{abstract}
\end{@twocolumnfalse}
]

\section{Introduction}
	
	The elements Ni and Pd play a key role in a wide range of catalytic applications. Functional materials and surfaces based on Pd are applied e.g. for heterogeneous catalysis and for hydrogen purification. For industrial applications, several aspects play an important role such as the chemical composition of the Pd surface, which influences the catalytic properties substantially, and the mechanical stability e.g. of thin membranes, which is affected by foreign atoms or segregation processes.
	For hydrogenation, Ni is the central catalyst for more than 80 years \cite{Adk30}. Since the invention of Raney \cite{Ran27} Ni as a catalyst is present in many synthetic transformations ranging from cross coupling reactions in which carbon-carbon bonds are formed to the reduction of electron rich carbon bonds.
	
	Nowadays, combinations of different catalytically active materials are used in order to increase efficiency and reduce costs. For example, del Rosario et al. \cite{Ros14} showed that the efficiency of direct-ethanol fuel cells can be increased by using a catalyst with a Ni-Pd bilayer structure instead of pure Pd. 
	Theoretical calculations made for Ni on a Pd surface predict Ni atoms to migrate in the second atomic layer \cite{Foi86, Lov05} for all crystallographic Pd orientations \cite{Boz03}. Experimental results on a \mbox{Ni-1$_\mathrm{at}$\,\% Pd} alloy studied with Auger Electron Spectroscopy (AES) in a temperature range of 550 to 800\,$^\circ$C show a moderate enrichment of Pd of the surface \cite{Mer79}. In Low Energy Electron Diffraction (LEED) studies on a Ni$_{50}$Pd$_{50}$(100) surface an oscillatory segregation profile with a Pd enriched surface layer and Ni enrichment in the second layer is observed \cite{Der95}.
	Khanra et al. demonstrated the effect of surface segregation on the catalytic activity of alloys using CO hydrogenation on a Pd–Ni(111) surface \cite{Kha98}. The influence of the annealing temperature to the surface segregation was shown by Li et al. on Au/Pd(111) using LEED and Low Energy Ion Scattering (LEIS) \cite{LiZ08}. 
	To achieve long-term stability of more sophisticated catalysts, the driving forces at surfaces leading to e.g. segregation, diffusion and temperature dependent interaction of surface atoms need to be understood, since they may significantly influence the catalytic activity. 
	
	The complex interplay of surface processes is subject of the presented first time in-situ temperature-dependent study on the stability of Ni adatoms on single crystalline Pd substrates with three different low index surface orientations. 
	In order to characterize a low Ni coverage, i.e. Ni with a nominal thickness of less than one monolayer (ML), on a Pd surface or even to observe surface segregation a method of highest surface sensitivity is required. For this reason we apply Positron annihilation induced Auger Electron Spectroscopy (PAES), which is known as an analysis method with outstanding surface sensitivity \cite{Meh90, Koy92a,May10c}. Auger electrons are emitted with an element-specific energy from ionized atoms in a radiation-free transition \cite{Aug25}. 
	Low energy positrons from the positron source NEPOMUC with an energy of less than 20\,eV are used to ionize an atom and induce the Auger process \cite{Hug07}. For this technique, the vast majority of the Auger electrons originate from the topmost atomic layer of the sample \cite{Faz04}, allowing the investigation of the elemental composition of the topmost surface layer. In addition, X-ray induced Photoelectron Spectroscopy (XPS) is applied to verify the comparability of the prepared samples, to analyze the elemental composition of the surface near layers, and to evaluate the influence of residual gas.

\section{Experimental setup and \\sample preparation}
	
	The measurements were carried out at the surface spectrometer \cite{Zim16} of the positron source NEPOMUC. The setup consists of several interconnected UHV-chambers for cleaning and preparation, analysis and storage of samples without exposing them to ambient conditions.  All samples were initially cleaned using Ar ion sputtering. Pd single crystals with the orientations Pd(111), Pd(110) and Pd(100) and a purity of 99.99\% were used as substrates. The Ni coating layer was prepared using an effusion cell equipped with a Ni rod with a purity of 99.99\,\%. For the PAES measurements, the energy of the positrons was set to 18\,eV. A hemispherical energy analyzer with CCD readout records the Auger electrons spectra. The sample surface is aligned perpendicular to the analyzer. The sample holder allows in-situ heating of the samples up to 500\,$^\circ$C by electron impact heating from the backside without producing significant background in the PAES spectra. The sample temperature is monitored by a thermocouple on the backside of the sample. It controls at the same time the heating power to keep the temperature constant. Additionally, the setup is equipped with an X-ray source with an Al anode, which is used for complementary XPS.

\section{Measurement and data \\evaluation}
	\label{sec:Measurement}
	
	\begin{figure}
		\centering
		\includegraphics[width=0.5\textwidth]{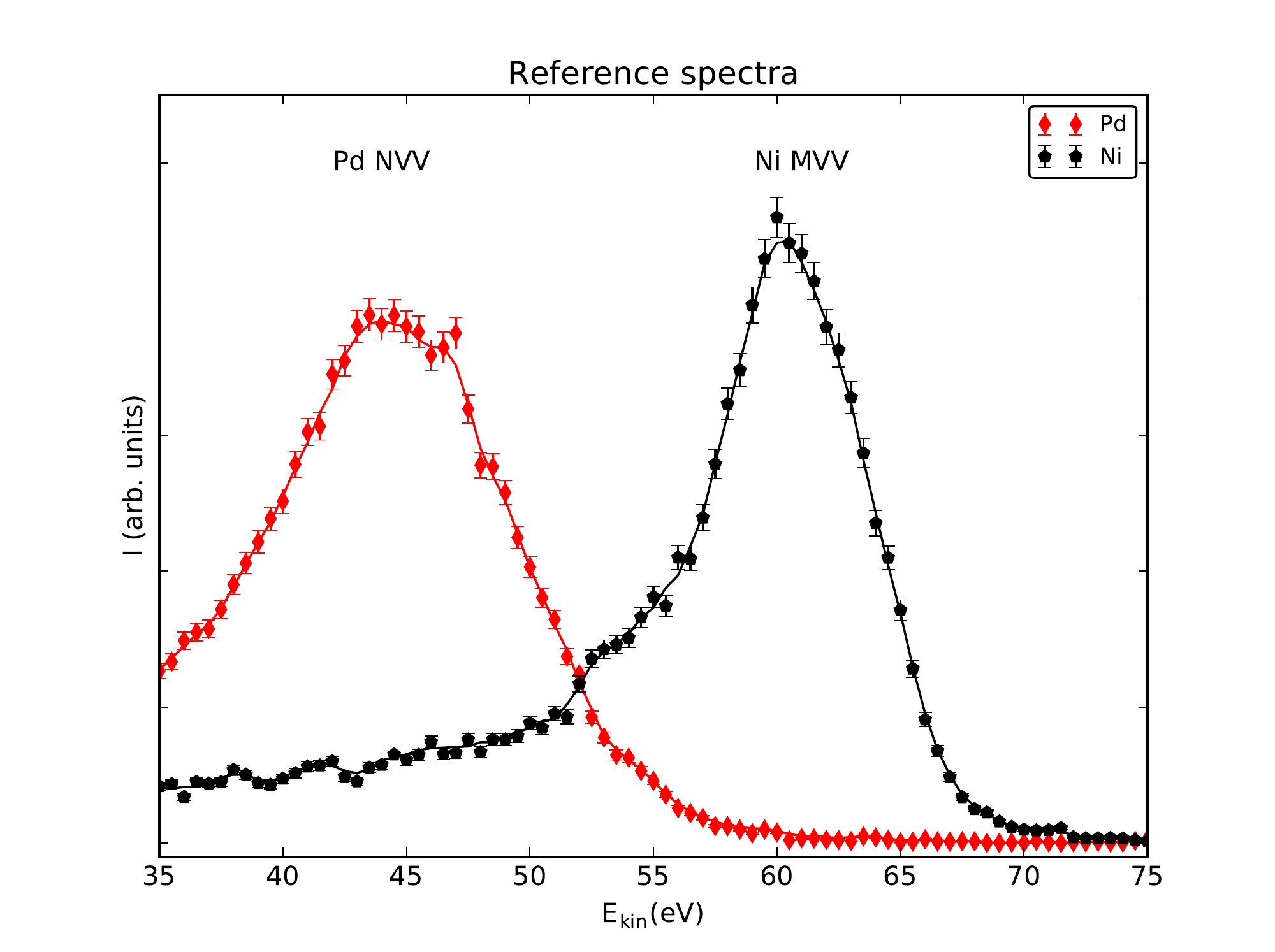}
		\caption{PAES reference spectra of Pd and Ni.}
		\label{fig:NiPd-RefSpectra}
	\end{figure}

	As a reference the PAES spectra of pure Pd and pure Ni were recorded in advance in an energy range of 35\,eV to 75\,eV showing the Pd NVV and Ni MVV Auger transitions (Figure \ref{fig:NiPd-RefSpectra}). The fixed retarding ratio-mode with a step width of 0.5\,eV and a retarding ratio of one was used. These detector parameters were also selected for the actual measurement on the Ni/Pd samples.
	The measurement time for a PAES spectrum amounted to 40 minutes. Each PAES measurement was accompanied by an XPS recording. The analysis of the photopeaks allows an estimation on the elemental composition in the near-surface layers. Moreover, surface contamination caused by the residual gas in the UHV chamber and chemical shifts can be detected. During the whole measurement time, the temperature of the sample is kept constant, before increasing by the next temperature step.

	To quantify the PAES spectra, the recorded data M(E) of a compound is fitted by a model function F(E), which is assumed as a linear combination of the reference spectra S$_{X,Y}$(E) of the pure elements X and Y using independently scaled fractions $\alpha$ and $\beta$:
	
	\begin{equation}
	F(E) = \alpha \cdot S_X(E)+\beta \cdot S_Y(E)
	\label{eq:FitFunction}
	\end{equation}
	As a figure of merit, the fit is characterized by the R value which quantifies the difference between the recorded data and the fit function. The R value as defined in \mbox{Equation \ref{eq:FitMinimize}} is minimized in the fitting procedure.
	
	\begin{equation}
	\mathrm{R}=\sum_i(\mathrm{M}_i-\mathrm{F}_i)^2
	\label{eq:FitMinimize}
	\end{equation}
	
	In the applied fitting procedure it is not a requirement that the sum of the scaling factors $\alpha$ and $\beta$ has to be unity. This condition would presuppose that the line shape of an Auger transition is not influenced by the presence of additional elements, or by chemical bonding. For this reason, the scaling factors are varied independently. 
	For the fitting procedure three steps of processing are performed: (i) a constant background represented by the minimum value in the spectrum is subtracted, (ii) each data point is weighted with its statistical error, and (iii) the spectrum area is normalized to one.
	
	For the evaluation of the XPS spectra the \mbox{Ni 2p} and \mbox{Pd 3d} photopeaks were quantified, using the standard quantization of CasaXPS \cite{CasaXPS}. A Shirley function representing the background is subtracted from the photopeak and the remaining peak area is quantified using the relative sensitivity factor of the transition.

\section{Results and discussion}
	
	\begin{figure}
			\centering
			\includegraphics[width=1.0\columnwidth]{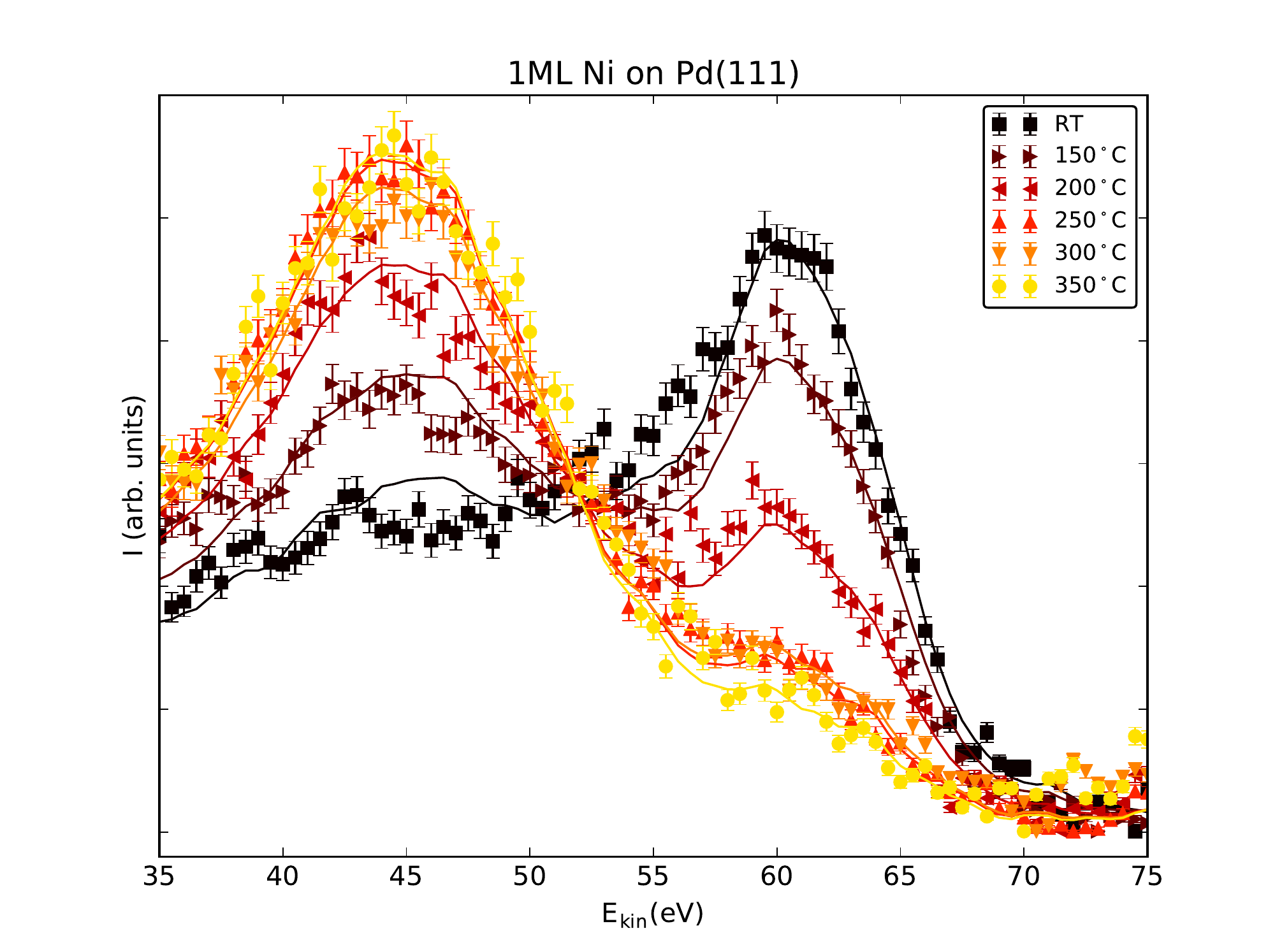}
			\label{fig:Pd111-Spec}
			\centering
			\includegraphics[width=1.0\columnwidth]{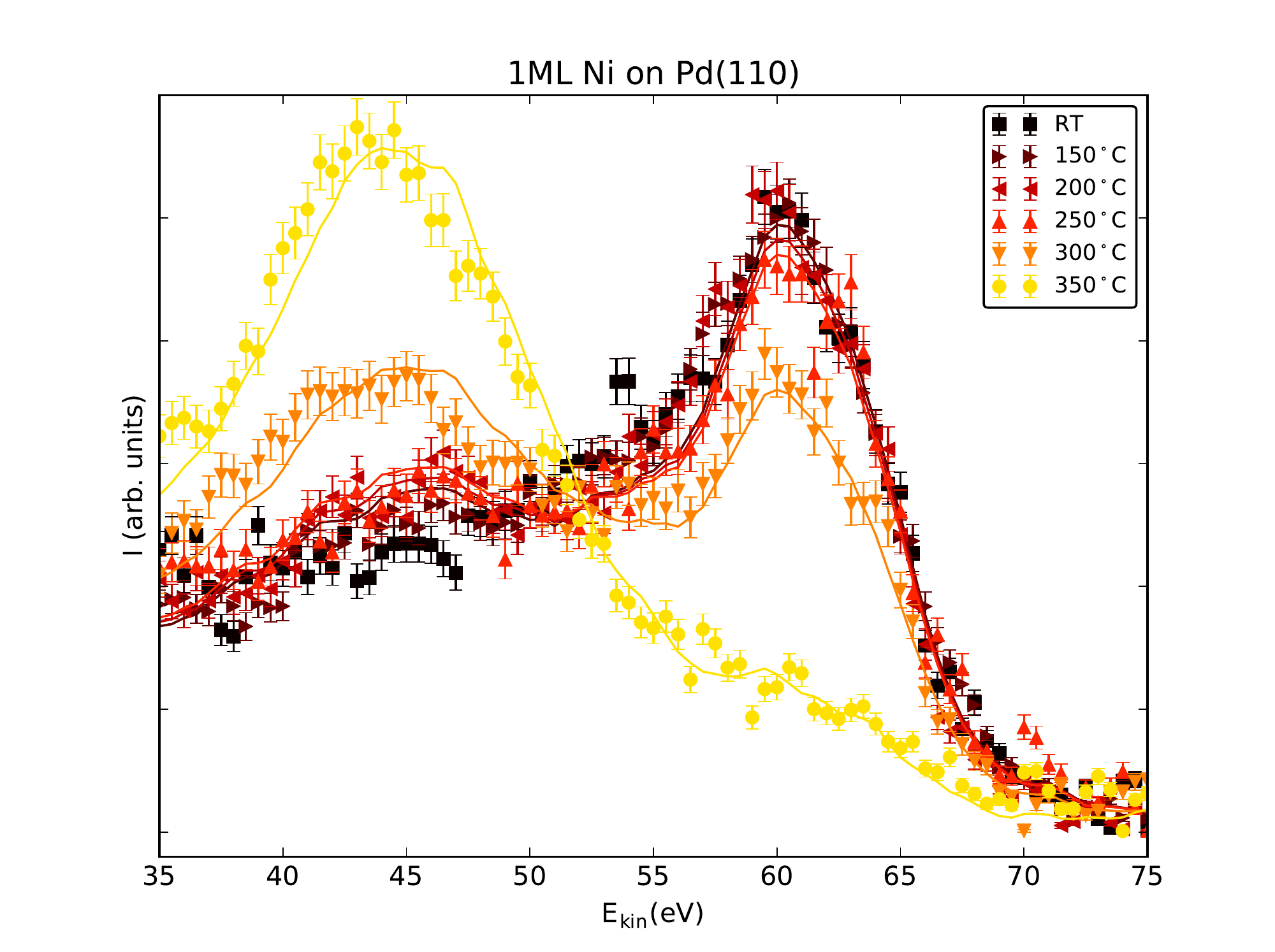}
			\label{fig:Pd110-Spec}
			\centering
			\includegraphics[width=1.0\columnwidth]{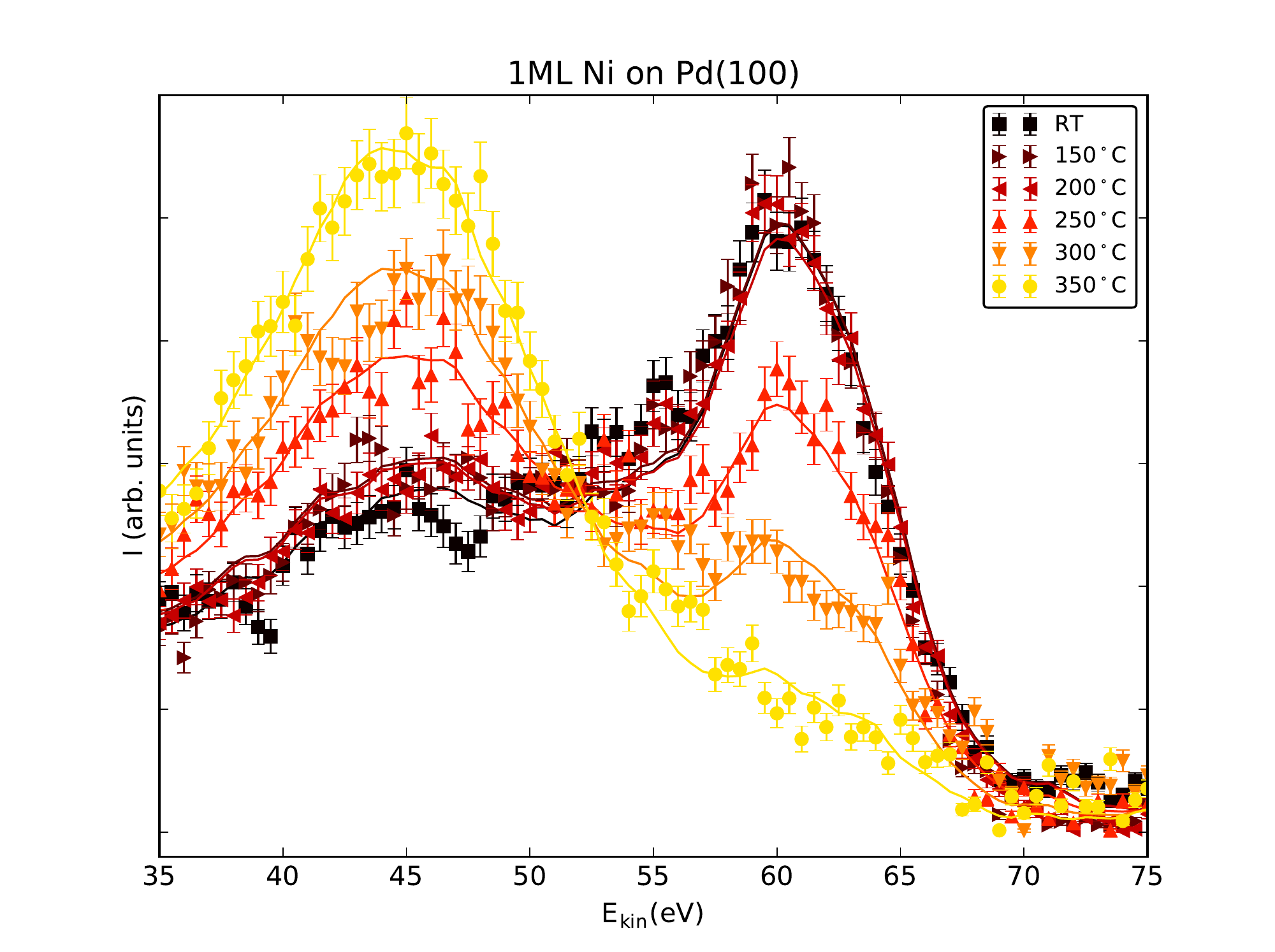}
			\label{fig:Pd100-Spec}
		\caption{Temperature dependent PAES spectra of one ML of Ni on Pd(111), Pd(110) and Pd(100). The solid lines represent the fit result of Equation \ref{eq:FitFunction} for the quantification of the Auger intensities of Ni and Pd, respectively.}
		\label{fig:Pd-Spectra}
	\end{figure}
	
	Figure \ref{fig:Pd-Spectra} shows the PAES spectra of the three different Pd substrate orientations coated with a single atomic layer of Ni in the temperature range between 30\,$^\circ$C and 350\,$^\circ$C. The Auger transitions Pd NVV at 43\,eV and Ni MVV at 64\,eV are clearly detectable. These low energy Auger transitions have a width of several eV, which mainly results from the broad valence band. Because of the isoelectronicity of Ni and Pd the effective atomic number and hence the Auger yield for the MVV and NVV transition, respectively, is equivalent. 
	
	Both, temperature and substrate orientation appear to strongly influence the relative intensities of the two elements. A fast decrease of the Ni intensity is clearly observed in the raw spectra at moderate temperatures of 150\,$^\circ$C for the Pd(111) substrate, whereas for Pd(110) no significant decrease can be detected up to 250\,$^\circ$C. Independent of the Pd substrate orientation, at 350\,$^\circ$C the spectrum is dominated by the Pd signal for all samples.
	
	\begin{figure}
		\begin{minipage}[t]{0.5\textwidth}
			\centering
			\includegraphics[width=1\textwidth]{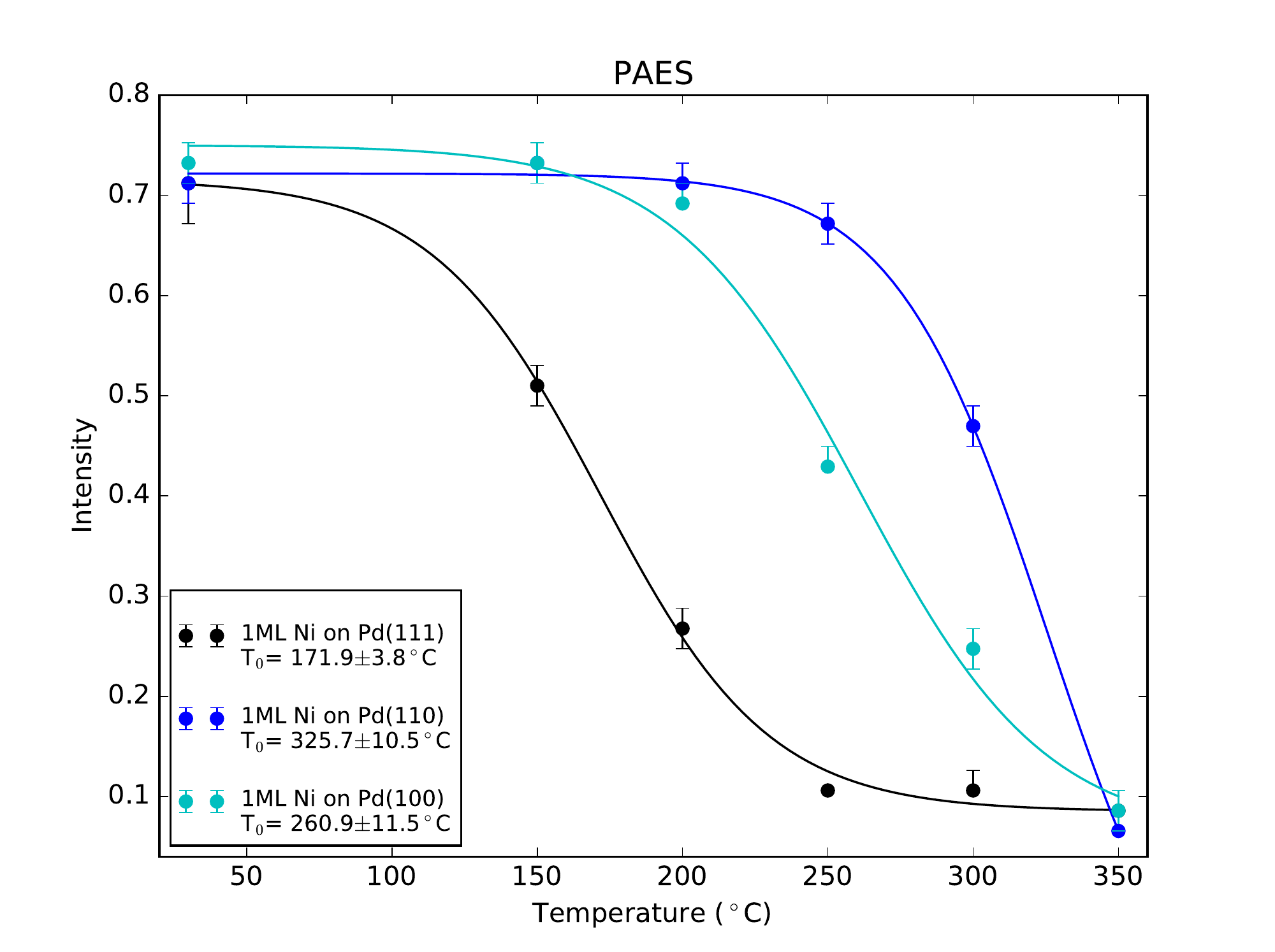}
			\label{fig:NiIntensityPAES}
		\end{minipage}
		\begin{minipage}[t]{0.49\textwidth}
			\centering
			(a)
		\end{minipage}
		\begin{minipage}[t]{0.5\textwidth}
			\centering
			\includegraphics[width=1\textwidth]{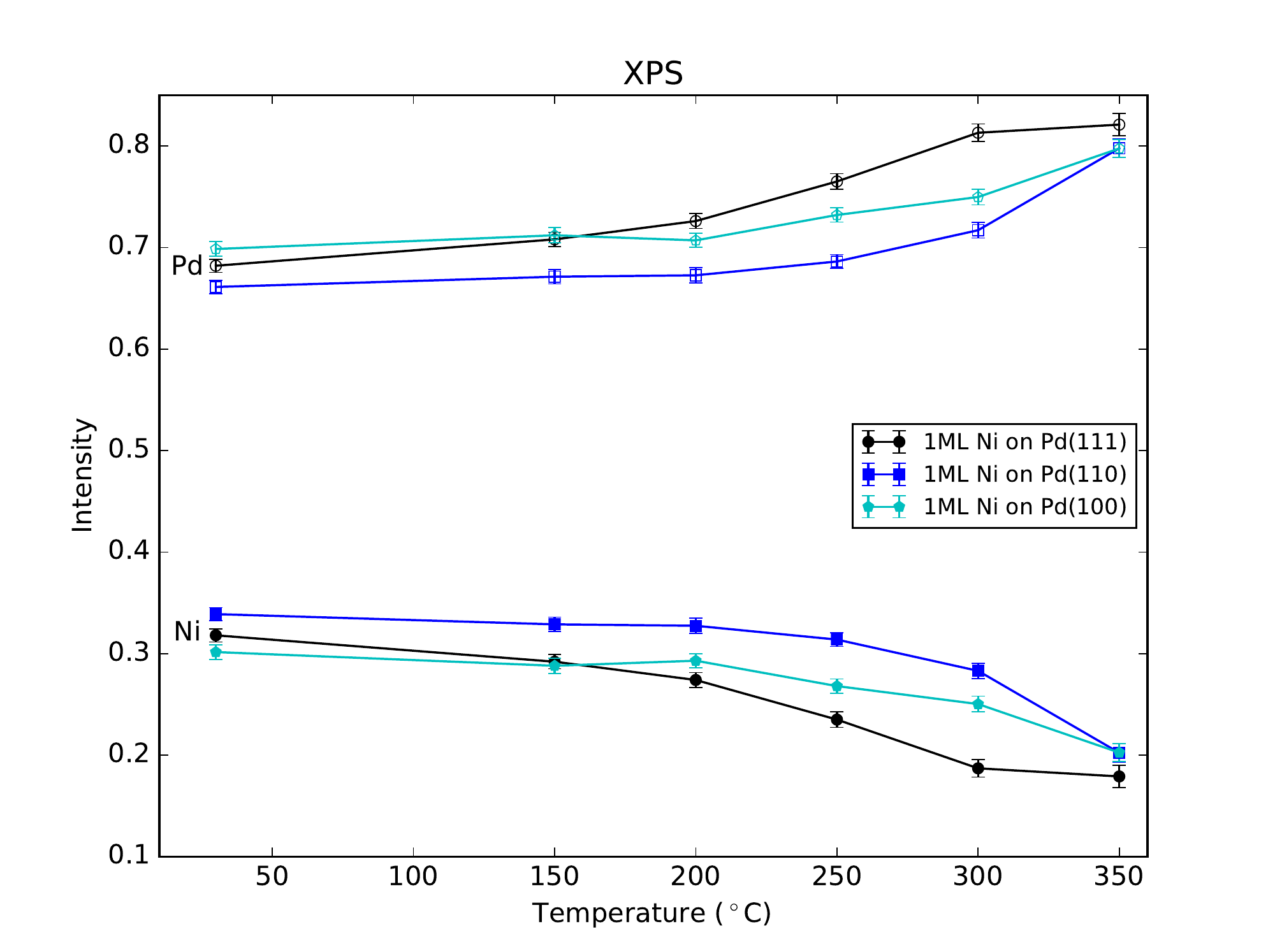}
			\label{fig:NiIntensityXPS}
		\end{minipage}
		\begin{minipage}[t]{0.5\textwidth}
			\centering
			(b)
		\end{minipage}	
		\caption{a) Ni intensity of the PAES spectra as a function of temperature for different Pd substrate orientations. The activation temperatures T$_0$ result from the fit function given in Equation \ref{eq:LogisticFunction}. b) Intensities for Ni and Pd derived from the XPS data. Note the same axis scale of the intensity.}
		\label{fig:NiIntensity}
	\end{figure}
	
	The measured data are well reproduced by the model function of the reference spectra discussed in Section \ref{sec:Measurement}. The fit results are shown as solid lines in the spectra of \mbox{Figure \ref{fig:Pd-Spectra}} and the Ni intensity is quantified in Figure \ref{fig:NiIntensity}a, where a value of one represents pure Ni.
	An average Ni intensity of 0.72 is detected at room temperature, but only 0.08 at 350\,$^\circ$C. Although a nominal thickness of one atomic layer of Ni was put on the Pd, the measurement signal also shows a small contribution of Pd. Despite a low evaporation rate of only 6\,\AA/minute, a certain inhomogeneity of the layer thickness can not be excluded. Moreover, a fraction of 5 to 20\,\% of the PAES signal may originate from the second atomic layer depending on the positron probability density at the surface \cite{Jen90}. 
	All Pd orientations show a strong tendency for segregation as predicted from theoretical calculations, leading to a Ni depletion of the surface. This corresponds to the fact that at a temperature of 350\,$^\circ$C, only a very low Ni amount can be detected for all Pd surfaces. The course of the Ni decrease can be described by a simple model, implying that an activation energy is needed to initiate the segregation process. The temperature dependency can hence be expressed by the equation 
	
	\begin{equation}
	I(T)=\frac{-C_1}{1+\exp(-C_2\cdot(T-T_0))}+I_0
	\label{eq:LogisticFunction}
	\end{equation}
	
	where T$_0$ denotes the corresponding activation temperature, while C$_1$ and C$_2$ are shaping parameters. The results of the quantification in Figure \ref{fig:NiIntensity}a were fitted using this model, leading to the lowest activation temperature of 172(4)\dC for Pd(111), followed by 261(12)\dC for Pd(100) and 326(11)\dC for Pd(110). The exact value and its error for each crystallographic orientation are given in Figure \ref{fig:NiIntensity}a.
	The corresponding activation energies amount to 38.4(3)\,meV for Pd(111), 46.0(10)\,meV for Pd(100) and 51.6(9)\,meV for Pd(110). 
	
	The complementary XPS data was evaluated by quantifying the photopeak intensity as described in the previous section, cf. Figure \ref{fig:NiIntensity}b. The mean detected Ni intensity in the three as-prepared samples at room temperature is 0.32. The individual samples show only slight deviations from the mean value and thus indicate a uniform sample preparation. No peak shifts of the Pd 3d and Ni 2p photo peaks were found in the data evaluation, which would indicate a contamination due to the rest gas during the measurement. A decrease of the Ni intensity with increasing temperature is found for all samples, showing a minimal value of 0.19 at 350\,$^\circ$C. This is a reduction of only 40\% compared with nearly 90\% for the PAES results. The XPS data substantiate the PAES results, though they are less significant because of the lower surface sensitivity of XPS in this energy range. However, the observed increase of the Pd intensity and the stabilization of the measured Ni signal at high temperatures support our interpretation of observing the surface segregation.

	The course of the Ni depletion depending on the Pd orientation is in good agreement with theoretical models for adatom surface diffusion on low index fcc surfaces, based on Monte Carlo simulations on Pd \cite{Agr02} as well as to those based on the embedded atom method calculated for Pt surfaces \cite{Liu91}. The exchange diffusion mechanism for the (100) orientation is taken into account. This mechanism is reported by Antczak and Ehrlich \cite{Ant07} and Oura et al. \cite{Our2003} to play an important role in surface diffusion for Pt, Ir and Ni and is predicted for Pd, too \cite{Liu91}.
	Regarding the low index surfaces, the lowest activation energy is reported for the (111)-surface, followed by the (100)-orientation and resulting in the highest value for (110) surface orientation. A variety of theoretical models and calculation schemes were compared, showing a uniform trend, but the exact results for the activation energies differ in a wide range between 0.14\,eV up to 1.35\,eV \cite{Agr02} and between 7\,meV and 4\,eV \cite{Liu91}. The experimentally derived energies are found in the range of some ten meV. 
	 
	The difference in the activation energy between the Pd(111) and Pd(110) orientation is less than a factor of two, compared with a factor of at least six for the theoretical calculations. The deviations may result from the fact that no segregation effects were included in the considered theoretical models, which may enforce the migration of Ni and lower the activation energies. The measured energies also indicate, that the energy difference for different orientations in the theoretical model may be overrated.   
	
	\section{Conclusion}
	
	In this paper we reported on the direct observation of surface segregation of Ni adatoms on single crystalline Pd surfaces, which for the first time was carried out in-situ temperature dependent. One single layer of Ni atoms was grown on Pd with the crystallographic orientations Pd(111), Pd(110) and Pd(100). The temperature dependent measurements of the composition of the first atomic layer have shown evidence for surface segregation effects related to crystal orientation. For all samples a strong tendency for Pd segregation was found, beginning at different activation temperatures T$_0$. 
	The lowest activation temperature of T$_0$=172\dC was evaluated for Pd(111), followed by Pd(100) with T$_0$=261\,$^\circ$C. The highest activation temperature was found for Pd(110) with T$_0$=326\,$^\circ$C. This sequence is in good agreement with theoretical predictions for adatom diffusion on low index fcc surfaces and substantiates the model of Pd surface enrichment for Ni/Pd due to surface segregation for all Pd orientations. However, the corresponding activation energies are on average by a factor of ten lower than predicted by theoretical models considering single atom diffusion and the exchange diffusion mechanism. More sophisticated models including full coverage by 1\,ML Ni of a Pd surface for the calculations of the temperature dependent surface kinetics are expected to improve the agreement between theory and experiment considerably.
	
	\section*{Acknowledgment}
	
	Financial support by the German Federal Ministry of Education and Research within the project no. BMBF-05K13WO1 is gratefully acknowledged.
	
	
	\bibliographystyle{unsrt}

\end{document}